\documentclass[12pt]{article}

\usepackage{amssymb}

\usepackage{graphicx,amssymb,url,mathrsfs, amsmath}
\usepackage{eucal}
\usepackage{wrapfig}
\usepackage{boxedminipage}
\usepackage{setspace}
\usepackage{subfigure}
\usepackage[dvipdfm]{hyperref}

\def\a  {\alpha}       \def\b  {\beta}         \def\g  {\gamma}
       \def\d  {\delta}        
\def\e  {\epsilon}        \def\k  {\kappa}

 \newcommand{\call}{\mbox{${\cal L}$}}

\def\IR{{\hbox{{\rm I}\kern-.2em\hbox{\rm R}}}}
\def\IB{{\hbox{{\rm I}\kern-.2em\hbox{\rm B}}}}
\def\IN{{\hbox{{\rm I}\kern-.2em\hbox{\rm N}}}}
\def\IC{\,\,{\hbox{{\rm I}\kern-.59em\hbox{\bf C}}}}
\def\IZ{{\hbox{{\rm Z}\kern-.4em\hbox{\rm Z}}}}
\def\IP{{\hbox{{\rm I}\kern-.2em\hbox{\rm P}}}}
\def\IH{{\hbox{{\rm I}\kern-.4em\hbox{\rm H}}}}
\def\ID{{\hbox{{\rm I}\kern-.2em\hbox{\rm D}}}}

\def\be{\begin{equation}}
\def\ee{\end{equation}}

\def\half{\frac{1}{2}}
\def\third{\frac{1}{3}}

\newcommand{\bra}[1]{\mbox{$\langle #1 |$}}
\newcommand{\ket}[1]{\mbox{$| #1 \rangle$}}

\def\tr{{\rm tr}\,}

\def\l{\left}
\def\r{\right}

\def\nn{\nonumber}

\def\dmu{\partial_{\mu}}
\def\dnu{\partial_{\nu}}

\newcommand{\calv}{{\cal V}}

\begin{document}

\begin{titlepage}

\begin{flushright}

\end{flushright}
\vspace{0.5in}

\begin{center}
{\large $\g^* \g^* \rightarrow \pi^0$ Form Factor from AdS/QCD\bf }\\
\vspace{10mm}
Alexander Stoffers and Ismail Zahed\\
\vspace{5mm}
{\it Department of Physics and Astronomy, Stony Brook University, Stony Brook NY 11794}\\
          \vspace{10mm}
{\tt \today}
\end{center}
\begin{abstract}
The recently measured $\gamma\gamma^*\rightarrow \pi^0$ anomalous form factor is analyzed using the $D4/D8\overline{D8}$ holographic  
approach to QCD. The half-on-shell transition form factor is vector meson dominated and is shown to exactly tie to the charged pion form 
factor. The holographic result compares well with the data for the lowest vector resonance. 
\end{abstract}
\end{titlepage}

\renewcommand{\thefootnote}{\arabic{footnote}}
\setcounter{footnote}{0}



{\bf 1.\,\,\,}Recently the BaBar collaboration has extended the measurement of the half-on-shell $\gamma\gamma^*\rightarrow \pi^0$  transition form factor up to $Q^2\approx 40\,{\rm GeV}^2$ photon virtualities~\cite{Aubert:2009mc,DRUZ}. The reported measurements  are considerably above the predicted  values using  factorization and  pQCD~\cite{Lepage:1979zb,Lepage:1980fj,Brodsky:1981rp}.  Although seen as a key benchmark for pQCD,  this exclusive process is tied with the flavor triangle anomaly in QCD and maybe more subtle. Similar difficulties were reported earlier by the JLAB collaboration for fixed angle Compton scattering $\gamma p\rightarrow \gamma p$ \cite{CEBAF}. 

A number of analyses have been put forward to try to reconcile the BaBar data with pQCD factorization through a modification of the pion distribution amplitude~\cite{RAD,POL, Dorokhov:2010bz}, whereby the pion distribution amplitude is argued to be flatter. However, there are difficulties in reconciling these modifications with the data at lower $Q^2$ which are seen to demand a vanishing pion distribution amplitude at the edges~\cite{STEF}.  

In this letter, we will put aside the idea of factorization and analyze the BaBar data using holographic QCD, a fully non-perturbative framework.  Our analysis will be based on the top-down dual construction~\cite{Sakai:2004cn,Sakai:2005yt}, in contrast to the bottom-up constructions recently discussed in~\cite{Grigoryan:2008up,Cappiello:2010uy,Brodsky:2010cq}. In the bottom-up approach ~\cite{Grigoryan:2008up} with a hard-wall the pion wave function needs an additional boundary term. As pointed out in \cite{Cappiello:2010uy}, the model studied here can be view as a hard-wall model albeit no changes to the pion wave function are necessary. While differences between these two approaches will show up in the IR of the boundary theory, we expect similarities in the UV. An interesting analysis within the context of large-$N_c$ Regge models is given in \cite{Ruiz Arriola:2006ii}.\\
\vspace{24pt} \\
{\bf 2.\,\,\,}
The $\pi^0 \g^* \g^*$ form factor can be assessed in holographic QCD using the $D4/D8\overline{D8}$ embedding  formulated by Sakai and Sugimoto~\cite{Sakai:2004cn, Sakai:2005yt} which supports vector meson dominance. Specifically ($k=q_1+q_2$ and $Q_{1,2}^2=-q_{1,2}^2$))
\be 
\int d^4x e^{-iq_1 x} \bra{\pi^0(k)} T \big(J^{\mu}_{em}(x) J^{\nu}_{em}(0)\big) \ket{0} = \e^{\mu \nu \a \b}q_{1\a}q_{2\b} F_{\g^*\g^* \pi^0}\l(Q_1^2,Q_2^2\r) \ , 
\label{defformfactor}
\ee
is saturated at tree level by vector meson resonances
\be
F_{\g^*\g^* \pi^0}\l(Q_1^2,Q_2^2\r) = \frac{N_c}{12 \pi^2 f_{\pi}} 
\sum_{m,n}\,a_{m} a_{n} c_{mn} \frac{1}{1+\frac{Q_1^2}{m_{m}^2}}\frac{1}{1+\frac{Q_2^2}{m_{n}^2}} \ ,
\label{formfactor}
\ee
where the $a_n$ characterize the vector couplings to the external EM current and the $c_{mn}$ the anomalous
$\pi^0$ coupling to the vectors (see \cite{Sakai:2004cn, Sakai:2005yt} and Appendix for details). In particular, the vector
couplings obey the sum rule
\be
\sum_{mn}a_ma_nc_{mn}=1 \ ,
\label{IDENT}
\ee
which shows that at the photon point (\ref{formfactor}) is fixed by the Abelian anomaly
\be
F_{\g\g \pi^0}\l(0,0\r)= \frac{N_c}{12 \pi^2 f_{\pi}} \ .
\ee
\vspace{24pt} \\
{\bf 3.\,\,\,}
For one photon on-mass shell, the transitional pion form factor is 
\be 
K\l(0,Q^2\r)\equiv 
\frac{12 \pi^2 f_{\pi}}{N_c} F_{\g\g^* \pi^0}\l(0,Q^2\r) = \sum_na_{n} g_{n\pi\pi} \frac{1}{1+\frac{Q^2}{m_{n}^2}} , \
\label{K}
\ee 
where we have used~\cite{Sakai:2004cn, Sakai:2005yt} 
\be
\sum_ma_{m} c_{mn}=g_{n\pi\pi}\ .
\ee
For $n=1$ we have $g_{1}=g_{\rho\pi\pi}\approx 6$, the standard rho-pi-pi coupling.
\begin{figure}[!htbp]
  \begin{center}
  \includegraphics[width=10cm]{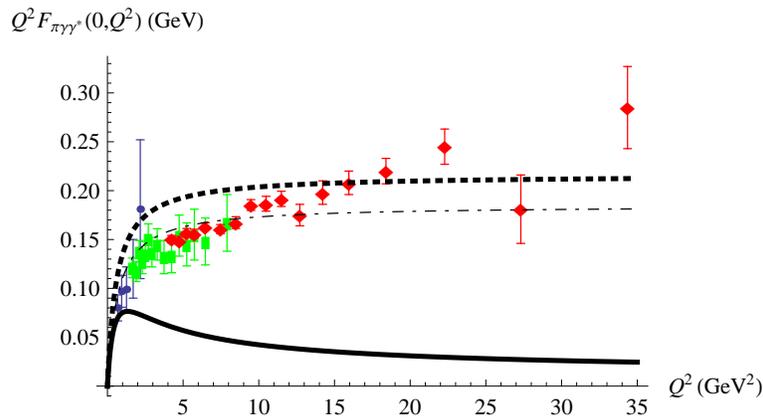}
    \caption{"(Color online)" Transitional pion form factor ($n=1$) vs. data. See text.}
  \label{figure02}
  \end{center}
\end{figure}
\begin{figure}[!htbp]
  \begin{center}
  \includegraphics[width=10cm]{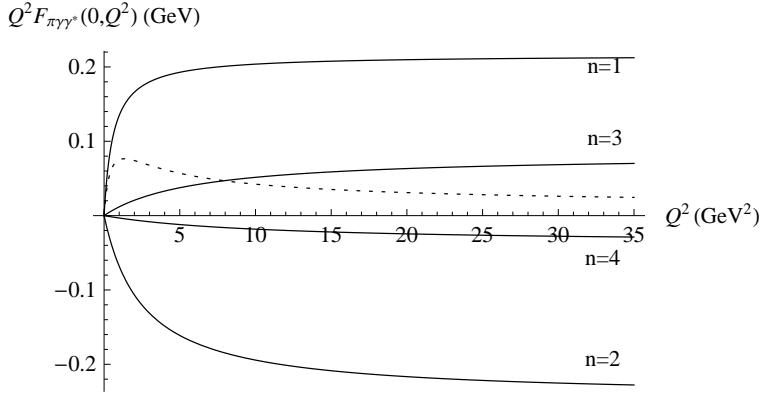}
  \caption{"(Color online)" Contributions to the transitional pion form factor. See text.}
  \label{figure02b}
  \end{center}
\end{figure}
In Fig.~\ref{figure02} we show the transitional pion form factor (for $n=1$) versus the data from Cello~\cite{Behrend:1990sr} (blue, circle), Cleo~\cite{Gronberg:1997fj} (green, square) and BaBar~\cite{Aubert:2009mc} (red, diamonds). We use $f_\pi= 0.0924$ GeV and $N_c=3$. The contribution from $n=1$ is shown with one photon on-shell (solid line) and one photon at $Q_1^2=0.18 \text{GeV}^2$ (dashed line). The dashed-dotted line is the pQCD interpolation~\cite{Brodsky:1981rp}
\be
Q^2 F_{\g\g^*\pi^0}^{\text{BL}}(0,Q^2)=\frac{Q^2}{4 \pi^2 f_\pi}\,\l(1+ \frac{Q^2}{8\pi^2f_\pi^2} \r)^{-1} 
\simeq \frac{Q^2}{4 \pi^2 f_\pi}\,\l( 1+ \frac{Q^2}{m_{\rho}^2}\r)^{-1} \ .
\label{BL}
\ee
The higher contributions from the holographic vectors are shown as the solid line contribution in Fig.~\ref{figure02b}. These vectors contribute with alternating sign to the transitional form factor and add up to zero asymptotically. The dotted line in Fig.~\ref{figure02b} shows the result for the transitional form factor including the first 8 resonances. Indeed \footnote{This can be checked by expanding the result in (\ref{K}) and using (\ref{wavefunctions}) as well as the completeness relation for the functions $\psi_{2n-1}$, see Appendix.}
\be
\lim_{Q^2 \rightarrow \infty} Q^2 K\l(0,Q^2\r) \simeq \sum_{n} a_{n} g_{n\pi\pi} m_{n}^2 = 0 \ . \label{largeQ}
\ee
As shown in \cite{Lichard:2010ap} the transitional form factor in a vector-meson-dominance model is sensitive to small $Q_1^2$. Here, the nature of the couplings dictated by the wave functions in the holographic direction yields a vanishing result for $Q^2 F_{\g\g^*\pi^0}$ at large $Q^2$ (independent of a non-vanishing $Q_1^2$), when the infinite tower of vector resonances is included. 
We recall that the top-down holographic approach effectively describes the QCD degrees of freedom for flavor excitations below $M_{KK}\approx 1\,{\rm GeV}$. When only the $n=1$ or rho resonance is retained, the large $Q^2$ asymptotic is
\be
\lim_{Q^2 \rightarrow \infty} Q^2 F_{\g\g^*\pi^0}(0,Q^2) \Big|_{n=1} \simeq  {a_1\,g_{1\pi\pi}} \frac{m_1^2}{4 \pi^2 f_\pi} = 1.31 \frac{m_1^2}{4 \pi^2 f_\pi} \  ,
\label{K1}
\ee
with $m_1=m_\rho$  and $a_1g_{1\pi\pi}\approx1.31$~\cite{Sakai:2005yt}.
This asymptotics, is in a better agreement with the data in the range $10 < Q^2 < 35 \ \text{GeV}^2$~\cite{Aubert:2009mc}. 
We recall that the pQCD result  does not vanish asymptotically~\cite{Lepage:1980fj}
\be
\lim_{Q^2 \rightarrow \infty} Q^2 F^{BL}_{\g\g^*\pi^0} = 2 f_\pi \simeq \frac{m_\rho ^2}{4 \pi^2 f_\pi} \ ,
\ee
where the last relation follows from the second KSRF relation $m_\rho^2=2g_{\rho\pi\pi}^2f_{\pi}^2$
with $g_{\rho\pi\pi}^2\approx 4\pi^2$.
The pQCD asymptotic (dashed-dotted line) is 30\% lower than the holographic asymptotic (dashed line) with
$n=1$ (rho meson only) as is explicit in Fig.~\ref{figure02}. 
\vspace{24pt} \\
{\bf 4. \,\,\,}
The charged pion form factor is studied in various holographic QCD models, see e.g. \cite{Grigoryan:2007wn, Kwee:2007dd, Kwee:2007nq, Brodsky:2007hb}. An analysis within large-$N_c$ Regge models is given in \cite{RuizArriola:2008sq}; see also \cite{Gorchtein:2011vf}. 
The model used here shows a rather unexpected result: For one photon on-mass shell, the transitional pion form factor is directly related to the 
charged pion form factor $F_\pi(Q^2)$ in holographic QCD:
\begin{figure}[!h]
  \begin{center}
   \includegraphics[width=10cm]{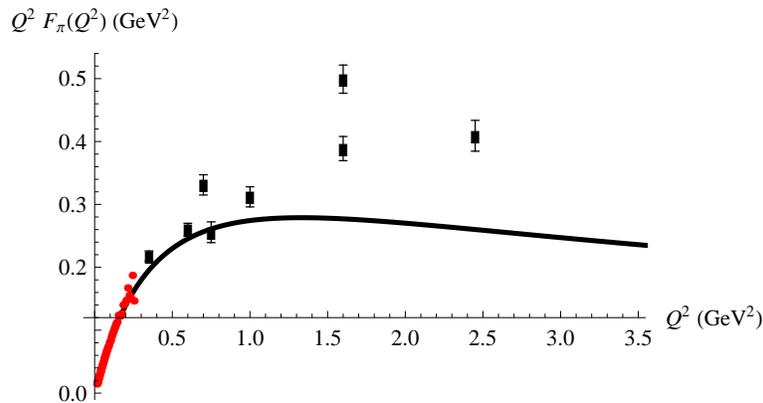}
   \caption{"(Color online)" Charged pion form factor from (\ref{KX}) with $n=1,...,8$. See text.}
   \label{figure01}
  \end{center}
\end{figure}
\be 
 F_{\pi}\l(Q^2 \r) =\sum_na_{n} g_{n\pi\pi} \frac{1}{1+\frac{Q^2}{m_{n}^2}} =K\l(0,Q^2\r) \ .
\label{KX}
\ee 
Note that the top-down model yields the same couplings for the charged and neutral pions. 
In Fig.~\ref{figure01} we show the behavior of the charged pion form factor following from (\ref{KX}) by using the first eight resonances ($n=1,...,8$). The data are from~\cite{Amendolia:1986wj} (red dots, error bars omitted for clarity) and from~\cite{Huber:2008id} (black squares). At small virtualities,
\be
K(0,Q^2)\approx 1-Q^2/m_1^2\approx 1-a_\pi\,Q^2/m_\pi^2 \ ,
\ee
where $a_\pi\approx 0.039$ can be tied to the pion charge radius by isospin
$a_\pi\equiv m_\pi^2\left<r^2\right>^{\pi}/6$. The measured
value is $a_\pi = 0.026 \pm 0.024 \pm 0.0048$~\cite{Farzanpay:1992pz}.
\begin{figure}[!h]
  \begin{center}
   \includegraphics[width=10cm]{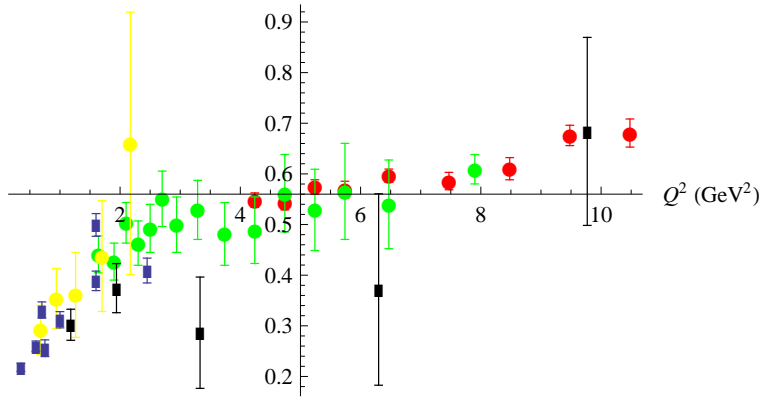}
   \caption{"(Color online)" Transitional form factor (circles) versus pion form factor (squares). See text.}
   \label{figure03}
  \end{center}
\end{figure}
The holographic relation (\ref{KX}) between the pion form factor and the transitional form factor implies a Ward-identity like relation at strong coupling. The consistency of this relation is checked in Fig.~\ref{figure03} where we have plotted the transitional form factor  $Q^2 4 \pi^2 f_\pi F_{\g\g^*\pi^0} \l(Q^2\r)$ from Cello 
(\cite{Behrend:1990sr}, magenta circles), Cleo (\cite{Gronberg:1997fj}, green circles) and BaBar \cite{Aubert:2009mc} (red, circles)  versus the measured pion form factor $Q^2 F_\pi \l(Q^2\r)$ from \cite{Bebek:1977pe} (black squares) and \cite{Huber:2008id} (blue squares) with $f_\pi = 0.0924$ GeV.  The latter data are only up to 10 ${\rm GeV}^2$. The identity is held rather well at low $Q^2$ and within the error bars at large $Q^2$.
\vspace{24pt} \\
{\bf 5. \,\,\,}
We have used the $D4/D8\overline{D8}$ holographic construction to analyze the pion transitional form factor. The transitional
form factor at large $N_c$ and strong coupling is entirely dominated by vector resonances while its on-shell
intercept is still fixed exactly by the Abelian anomaly. A comparison to the existing BaBar data implies that only
the $n=1$ or $\rho$ resonance should be retained to accomodate the measured data up to $Q^2=40$ 
${\rm GeV}^2$. This is consistent with the expectation that the $D4/D8\overline{D8}$ holographic model with vector excitations
works at or below the $M_{KK}\approx 1\,{\rm GeV}$ scale. The holographic construction ties the transitional pion form factor to the charged pion form factor.
This Ward-like identity is found to be well obeyed by the existing data for both form factors, including the recent
BaBar data up to $Q^2\approx 10$ ${\rm GeV}^2$.
\vspace{24pt} \\
{\bf Acknowledgements\,\,\,}
This work was supported in part by the  US Department of Energy under contract No. DE-FG-88ER40388.

\newpage

{\bf Appendix: Holographic Summary.\,\,\,}
\\
\\
In this Appendix we briefly note some of the holographic conventions and results of the $D4/D8\overline{D8}$ construction
of relevance to our analysis in the text. We refer to~\cite{Sakai:2004cn,Sakai:2005yt}  for further details. The effective Lagrangian 
(DBI plus Chern-Simons) contributions below the $M_{KK}$ scale are

\begin{eqnarray} \nn
\call_{\rm{D8}}^{\rm{DBI}}+\call_{\rm{D8}}^{\rm{CS}}&\approx&\half \tr \l(\dmu v_{\nu}^n-\dnu v_{\mu}^n \r)^2 + a_{n} \tr \l(\partial^{\mu} \calv^{\nu}-\partial^{\nu}\calv^{\mu}\r)\l(\dmu v_{\nu}^n -\dnu v_{\mu}^n\r) + m_{n}^2 \tr \l(v_{\mu}^n\r)^2 \\  
&- &\frac{i N_c}{4 \pi^2 f_{\pi}} \e^{\a \b \g \d}\tr \l(\Pi \partial_\a v_\b^n \partial_\g v_\d^m \r) c_{nm} \label{lagrangian}
\end{eqnarray}
with $U(N_f)$ valued pion ($\Pi$), photon ($\calv_\mu$) and vector ($v_\mu^n$) fields.
The vectors $v_\mu^n=i T^a v_\mu^{na}$ are $U(N_f)$ valued with the normalization 
$\tr \l(T^a T^b \r)= \d^{ab}/2$. Here and in the text the sum over the vector modes $m,n=1,2,3,....$
is implied. All the vector couplings in (\ref{lagrangian}) are fixed by the behavior of the holographic wave functions. Specifically, 

\begin{eqnarray} \nn
a_{n}=\k \int dz \ K^{-1/3} \psi_{2n-1} \ , \ \ c_{nm} = \frac{1}{\pi} \int dz \ K^{-1} \psi_{2n-1} \psi_{2m-1} \ .
\end{eqnarray}
with $K=1+z^2$.
$\k={\lambda N_c}/{216 \pi^3}\simeq 0.00745$ is fixed by the pion decay constant. 
The holographic wave functions $\psi_{2n-1}$ and the masses for the vector modes satisfy the equation

\begin{eqnarray}
-K^{-\third}\partial_z \l( K \partial_z \psi_{2n-1}\r) = \lambda_n \psi_{2n-1} \ , \ \ \ \lim_{z\rightarrow \pm\infty} \psi_{2n-1} \rightarrow 0 \ , \ \ \  \partial_z \psi_{2n-1}(0)=0 \ , \ \ \ m_{n}^2=\lambda_n M_{\rm{KK}}^2 \ . \label{wavefunctions}
\end{eqnarray}  
They are normalized by

\be
\k \int dz \ K^{-\third} \psi_{2n-1} \psi_{2m-1} = \d_{nm} \ .
\ee 
The scale of the vector masses is set by $M_{KK}\approx 1\,{\rm GeV}$.

\end{document}